# GreenPod: Energy-Optimized Scheduling for AIoT Workloads Using TOPSIS

Preethika Pradeep, Eyhab Al-Masri
School of Engineering and Technology
University of Washington
Tacoma, WA, USA
(ppreejit, ealmasri@uw.edu)

*Abstract*—**AIoT workloads demand energy-efficient orchestration across cloud-edge infrastructures, but Kubernetes' default scheduler lacks multi-criteria optimization for heterogeneous environments. This paper presents GreenPod, a TOPSIS-based scheduler optimizing pod placement based on execution time, energy consumption, processing core, memory availability, and resource balance. Tested on a heterogeneous Google Kubernetes cluster, GreenPod improves energy efficiency by up to 39.1% over the default Kubernetes (K8s) scheduler, particularly with energy-centric weighting schemes. Medium-complexity workloads showed the highest energy savings, despite slight scheduling latency. GreenPod effectively balances sustainability and performance for AIoT applications.**

## I. Introduction

The integration of Artificial Intelligence (AI) and the Internet of Things (IoT)—collectively termed AIoT [1]—is transforming modern computing by enabling distributed systems to process real-time data at scale. These intelligent systems are prevalent in smart cities, manufacturing, and healthcare, where energy efficiency, low latency, and adaptability are essential. AIoT workloads typically span heterogeneous cloud-edge infrastructures, requiring a balance between performance and energy consumption.

Furthermore, the adoption of the microservices architecture and containerization has revolutionized software development, promoting scalability and maintainability [2]. In parallel, serverless computing has accelerated the adoption of stateless, event-driven architectures, particularly within AIoT ecosystems where services react to sensor data streams [3, 4]. While service meshes help manage microservice interactions, critical decisions on placement and scalability are generally handled by an underlying orchestration layer—typically Kubernetes (K8s).

Kubernetes has become a de facto standard for container orchestration [5], with its scheduler determining where and when containerized microservices are executed. However, the default Kubernetes scheduler has some shortcomings because it primarily considers basic resource availability during task execution, which is inadequate in heterogeneous environments where multiple factors—such as energy efficiency, execution performance, and resource balance—should guide decisions [6]. This limitation becomes increasingly critical as data centers, a primary AIoT computing backbone, now consume about 1% of global electricity, projected to rise to 3–13% by 2030 [7, 8].

While traditional resource scheduling research has predominantly focused on virtual machines (VMs) [9], containers differ significantly due to their lightweight, dynamic nature. They often represent fine-grained workloads in edge computing environments, requiring faster and more adaptive scheduling decisions. This challenge is amplified in AIoT systems that are characterized by highly variable, energy-sensitive workloads [10]. Although some theoretical frameworks have proposed Multi-Criteria Decision Analysis (MCDA) methods for cloud-based resource scheduling, their integration into practical Kubernetes environments remains very limited [11, 12]. Moreover, few studies address the real-world complexities of IoT workloads and dynamic resource contention within Kubernetes clusters [16, 17].

This paper introduces GreenPod, an energy-aware Kubernetes scheduler that leverages the Technique for Order Preference by Similarity to Ideal Solution (TOPSIS)—an MCDA approach well-suited for environments with competing objectives. TOPSIS is particularly effective for Kubernetes scheduling as it systematically ranks pod placement options based on multiple weighted criteria. GreenPod optimizes pod placement by evaluating five key metrics: execution time, energy consumption, core availability, memory availability, and resource balance. To this extent, our key contributions throughout this study include the following:

- We developed GreenPod, an innovative Kubernetes-compatible scheduler that leverages the TOPSIS multi-criteria decision-making approach for intelligent and energy-efficient pod placement suitable for AIoT workloads.

- GreenPod incorporates weighted metrics such as execution time, energy use, and resource availability, addressing the default Kubernetes scheduler's single-objective limitations to enable more sustainable, adaptive workload management.

- We conduct a comprehensive evaluation on a heterogeneous Google Kubernetes Engine (GKE) cluster, demonstrating that GreenPod can achieve energy savings of up to 39.1%, particularly when employing energy-centric configurations.

- GreenPod significantly reduces carbon emissions (~3.39 metric tons per cluster annually) and enables economic gains via carbon credits ranging from $1.84 to $667, emphasizing its role in sustainable, cost-effective AIoT management.

- To support reproducibility and foster future research, we publicly release the GreenPod scheduler as an open-source tool, allowing researchers to integrate and extend it within Kubernetes-based AIoT environments [39].

By incorporating multi-criteria decision analysis within Kubernetes for scheduling tasks, GreenPod addresses the need for having a sustainable and an adaptive orchestration for next-generation AIoT systems. Its practical implementation bridges the gap between theoretical MCDA methods and real-world containerized environments, fostering more energy-efficient AI-driven IoT ecosystems.

## II. RELATED WORK

### A. Container Orchestration and Scheduling Mechanisms

The proliferation of containerization has sparked extensive research into efficient orchestration of distributed workloads. Kubernetes has emerged as a leading container orchestration platform, valued for its modular architecture and declarative management model [13]. Its default scheduler, however, primarily operates using a simple scoring-based system that mainly evaluates available CPU and memory resources [14, 15]. While this heuristic-driven approach suffices for basic workloads, it struggles in heterogeneous environments where multiple, competing objectives—such as energy consumption and execution latency—must be simultaneously addressed.

Wang et al. [16] highlighted the limitations of standard Kubernetes schedulers in IoT applications with stringent requirements in edge environments. Their Edge Information-Aware Scheduler achieved 18% lower latency and a 140% improvement in computing performance by incorporating network topology awareness. Similarly, Rossi et al. [17] noted that the Kubernetes native scheduler lacks network-awareness, which is crucial for latency-sensitive applications, particularly in geo-distributed microservices deployments. Their analysis revealed that default policies simply distribute containers across available cluster resources without considering network delays [17]—a significant shortcoming in cloud-edge infrastructures.

### B. Multi-Criteria Decision Analysis in Scheduling

To overcome Kubernetes single-objective limitations, researchers have proposed multi-criteria scheduling approaches. One notable example is the Kubernetes Container Scheduling Strategy (KCSS), which utilizes the TOPSIS algorithm for multi-criteria node selection [18]. KCSS evaluates nodes based on six factors: CPU utilization, memory utilization, disk usage, power consumption, running containers, and image transmission time. However, KCSS lacks a comprehensive analysis of weighting schemes and fails to adequately adapt to workload diversity and resource competition.

Furthermore, a limited number of research efforts have explored the use of Multi-Criteria Decision Analysis (MCDA) methods for Kubernetes scheduling. Among these, TOPSIS has been utilized to optimize container scheduling by allowing user-defined weighting of system parameters [37, 38]. However, most implementations are limited to homogeneous clusters and synthetic workloads, offering little insight into diverse, real-world scenarios [19]. Some other approaches combine multiple MCDA techniques, such as SAW, VIKOR, and COPRAS, but these approaches remain largely theoretical or confined to simulations, rather than practical, container-native environments or real-world deployments [20, 21]. Consequently, despite the potential of MCDA methods to enhance Kubernetes scheduling, their application in heterogeneous environments has remained limited—an issue we address by introducing GreenPod.

### C. Energy-Efficient Scheduling Approaches

The increasing carbon footprint of data centers has driven significant research into energy-efficient orchestration [22]. Zhang et al. [23] proposed a power-aware VM scheduler that demonstrated notable energy savings. However, VM-level orchestration differs from container scheduling in K8s, where granularity, volatility, and resource sharing are crucial [24].

Within Kubernetes, several strategies have emerged to optimize resource utilization and performance. Song et al. introduced the Gaia Scheduler, which focuses on distributing GPU loads by treating GPU resources similarly to CPUs [26]. While effective in GPU-centric scenarios, the Gaia Scheduler lacks a multi-criteria model applicable to diverse AIoT workloads [26]. Another notable approach is the faasHouse scheduler by Aslanpour et al., which employs computational offloading for energy-aware scheduling [27]. Despite its innovative use of resource sharing, faasHouse often overlooks execution time and system balance, leading to potential performance trade-offs [27].

Additionally, Piraghaj et al. proposed an energy-efficient container consolidation framework on virtual machines to minimize power consumption within cloud environments [28]. Santos et al. developed a MILP-based framework for resource provisioning in fog computing, optimizing IoT service allocation while balancing cloud and network requirements in smart city scenarios [29]. Piontek et al. introduced a $CO_2$-aware workload scheduling algorithm for K8s, leveraging historical data to schedule non-critical jobs and reduce $CO_2$ output [30]. However, Piontek's approach mainly targets batch processing, limiting its applicability in real-time IoT environments where adaptive, multi-criteria optimization is crucial.

### D. Limitations and Research Gap

Although container scheduling has been extensively studied, most approaches remain theoretical or focus on cost reduction rather than energy efficiency. In addition, many of the existing research methods target VM scheduling rather than container-native strategies, and even when MCDA methods are applied, they are rarely integrated into Kubernetes in a practical, deployment-ready mechanism. Furthermore, existing energy-aware schedulers often lack multi-criteria optimization or fail to address the diversity of real-world AIoT workloads [24].

To address these gaps, we introduce GreenPod—a TOPSIS-based custom Kubernetes scheduler designed for energy-efficient, multi-criteria optimization of IoT workload orchestration in AIoT environments. Unlike the default scheduler, GreenPod integrates critical metrics such as processor utilization, memory availability, energy consumption, and execution time into its adaptive decision-making process. Using TOPSIS for node ranking, it enables context-aware, sustainability-driven scheduling. GreenPod is well-suited for dynamic, resource-constrained systems requiring efficient task execution, real-time responsiveness, and enhanced operational longevity.

## III. SYSTEM ARCHITECTURE AND METHODOLOGY

GreenPod implements an energy-optimized orchestration framework for AIoT workloads through a hierarchical, multi-tier architecture. It leverages Kubernetes extensibility while incorporating energy-aware decision-making. As shown in Figure 1, the system consists of three primary tiers: heterogeneous edge devices, an intelligent edge gateway hosting the TOPSIS-based scheduler within the Kubernetes (K8s) environment, and an elastic cloud environment for workload offloading. This design ensures workload-aware placement that balances performance and energy efficiency—critical for sustainable AIoT operations.

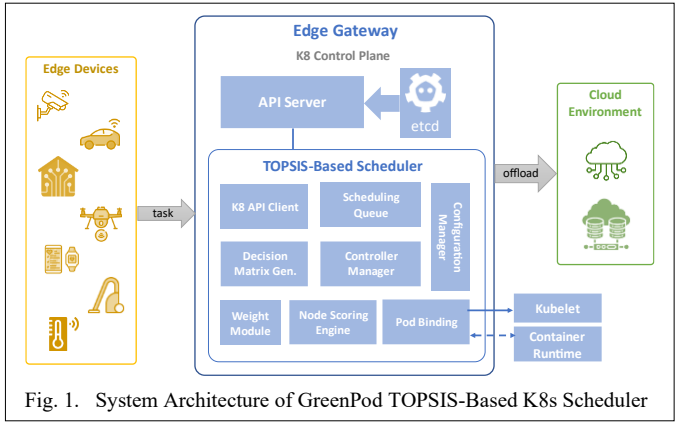

Fig. 1. System Architecture of GreenPod TOPSIS-Based K8s Scheduler

### A. GreenPod System Components

- **Edge Devices**: These IoT endpoints generate diverse computational loads and encapsulate domain-specific tasks into containerized workloads with distinct resource profiles. Workloads are submitted via *kubectl*, or APIs for automated deployment.

- **Edge Gateway**: This gateway hosts a Kubernetes cluster, worker services, and the GreenPod TOPSIS-based scheduler, serving as both a computational platform and an intelligent orchestrator.

- **K8s Control Plane**: In addition to standard Kubernetes components, GreenPod introduces monitoring agents that collect fine-grained energy data via hardware interfaces or calibrated power models. A secondary scheduler operates alongside the default *kube-scheduler*.

- **TOPSIS-Based Scheduler**: The core of GreenPod's architecture is its TOPSIS-based scheduler, designed to optimize pod placement through a multi-stage decision pipeline. The process starts with the energy profiling module, which monitors and predicts workload-specific energy consumption. An adaptive weighting module dynamically adjusts criteria weights based on system conditions. Then, a decision matrix generator constructs normalized matrices representing node performance metrics. The TOPSIS node scoring engine calculates distance measures and closeness coefficients to rank nodes. Once an optimal node is selected, the pod binding module binds the pod through the Kubernetes API server. A logging and monitoring component records outcomes and performance metrics for iterative improvements. This multi-stage pipeline ensures GreenPod balances energy efficiency and performance, making it suitable for dynamic workloads.

- **Cloud Environment**: The cloud acts as an offloading extension, utilizing workload classification to determine optimal placement. It also supports cross-environment synchronization and federated resource pooling to enable workload migration based on energy efficiency thresholds.

### B. Scheduler Implementation & Operational Flow

GreenPod enhances Kubernetes scheduling by integrating a custom scoring pipeline with energy-based filtering and multi-stage scoring to reduce computational overhead. Deployed as an additional component, it remains compatible with the default scheduler, enabling incremental adoption and fallback. It utilizes standard Kubernetes resources with scoped RBAC policies for secure operation.

Further, GreenPod seamlessly integrates into the Kubernetes scheduling lifecycle with energy-aware decision points. AIoT workloads, annotated with energy metrics, are intercepted by the scheduler, which evaluates nodes using TOPSIS-based scoring across performance and energy dimensions. The optimal node is selected based on the highest closeness coefficient, and energy consumption is continuously monitored to refine scheduling decisions.

## IV. EXPERIMENTAL DESIGN & IMPLEMENTATION

To assess the performance of our GreenPod TOPSIS-based scheduler, we developed an experimental framework that models real-world AIoT deployment scenarios and workloads. The methodology was designed to evaluate the scheduler's ability to dynamically manage microservices within heterogeneous edge computing environments.

### A. Kubernetes Cluster Environment

Our experiments were conducted on a running Google Kubernetes Engine (GKE) cluster configured to replicate the resource constraints and operational characteristics typical of edge computing running or executing IoT applications. The cluster featured a heterogeneous node setup to simulate realistic placement challenges, including variations in processing power, memory availability, and energy efficiency, reflecting conditions commonly found in production IoT deployments. Table I presents the employed K8s cluster configuration.

TABLE I. CLUSTER CONFIGURATION

| Node Category | Instance Type | vCPUs | Memory | Purpose |
|---|---|---|---|---|
| A | e2-medium (2 vCPU) | 2 | 4GB | Energy-efficient, minimal resources |
| B | n2-standard-2 (2 vCPU) | 2 | 8GB | Balanced performance |
| C | n2-standard-4 (4 vCPU) | 4 | 16GB | High-performance, high resource |
| Default | e2-standard-2 (2 vCPU) | 2 | 8GB | System components |

## B. Workload Characterization

We deployed a set of containerized workloads representing typical IoT data processing tasks commonly executed within edge environments. The selected tasks, summarized in Table II, emulate representative IoT applications such as anomaly detection, object detection, privacy preservation, and predictive maintenance, reflecting the varied computational requirements, resource demands, and operational complexity typical of edge computing deployments.

TABLE II. SUMMARY OF CONTAINERIZED WORKLOADS

| Workload Type | Description | Resource Requests | Task Size |
|---|---|---|---|
| Light | Basic Linear Regression → 1,000 samples | 0.2 CPU, 0.5GB | Small |
| Medium | Scalable Linear Regression → 1 million samples | 0.5 CPU, 1GB | Scalable |
| Complex | Distributed Linear Regression → 10 million samples | 1.0 CPU, 2GB | Distributed |

## C. Experimental Setup

The experimental methodology employed a structured factorial design to systematically evaluate the scheduler's performance under varying operational conditions. Key factors included competition levels (*low, medium, high*) and weighting schemes (*general, energy-centric, performance-centric, resource-efficient*), as shown in Table III.

TABLE III. EXPERIMENTAL SETUP FOR SCHEDULER EVALUATION

| Factor | Levels |
|---|---|
| Competition Level | Low, Medium, High |
| Weighting Scheme | General (Balanced), Energy-Centric, Performance-Centric, Resource-Efficient |
| Scheduler Type | TOPSIS, Default Kubernetes |

The evaluation protocol used multiple metrics to assess scheduling efficiency. Energy consumption (kJ) measures power usage from scheduling decisions, while scheduling time (ms) quantifies the algorithm's computational overhead. Node allocation efficiency tracks workload distribution as the ratio of actual to optimal allocation, indicating load balancing. Execution performance (s) evaluates system responsiveness and latency, reflecting service quality under varying conditions. The factorial design in Table IV evaluates combinations of competition levels and weighting schemes relevant to AIoT deployments.

TABLE IV. METRICS FOR ASSESSING SCHEDULING EFFICIENCY

| Metric | Description |
|---|---|
| Energy Consumed (kJ) | Quantifies the efficiency of scheduling decisions from an energy optimization perspective |
| Scheduling Time (ms) | Measures the computational overhead introduced by the scheduling algorithm |
| Node Allocation | Analyzes the distribution patterns and resource allocation preferences across the cluster |
| Execution Performance | Assesses workload-specific performance characteristics and system responsiveness |

## D. Weighting Schemes (Scheduling Profiles)

To accommodate different operational priorities, we implemented five weighting schemes, each tailored to a specific performance goal. The general scheme assigns equal importance to all metrics, providing a balanced evaluation. The energy-centric scheme prioritizes power consumption, optimizing for energy efficiency in resource-constrained environments. The performance-centric scheme emphasizes execution speed, suitable for latency-sensitive applications. Finally, the resource-efficient scheme balances overall resource utilization and energy efficiency, aiming to optimize performance without excessive power consumption. These schemes enhance the scheduler's adaptability to diverse requirements in heterogeneous edge environments.

## E. Competition Level Configuration

We established three competition levels to evaluate the scheduler's performance under different resource contention scenarios, as outlined in Table V.

TABLE V. COMPETITION LEVEL CONFIGURATION

| Level | Light Pods | Medium Pods | Complex Pods |
|---|---|---|---|
| Low | 4 (2 TOPSIS, 2 Default) | 2 (1 TOPSIS, 1 Default) | 2 (1 TOPSIS, 1 Default) |
| Medium | 8 (4 TOPSIS, 4 Default) | 4 (2 TOPSIS, 2 Default) | 2 (1 TOPSIS, 1 Default) |
| High | 12 (6 TOPSIS, 6 Default) | 6 (3 TOPSIS, 3 Default) | 4 (2 TOPSIS, 2 Default) |

These competition levels were designed to evaluate scheduler performance under varying resource contention scenarios. Low competition represents minimal contention, where resources are readily available. Medium competition simulates moderate demand with partial system utilization, balancing availability and contention. High competition reflects intensive resource contention, with near-full utilization requiring precise scheduling to maintain performance and energy efficiency.

## V. EVALUATION AND RESULTS

We evaluated the energy-centric scheduling solution against the default Kubernetes scheduler, focusing on energy efficiency across four scheduling profiles (*general or balanced, energy-centric, performance-centric, resource-efficient*) and three competition levels (*low, medium, high*).

### A. Experimental Results Overview

Energy-centric strategies consistently achieved the highest energy savings, particularly in low and medium competition scenarios, highlighting the benefit of prioritizing energy optimization. The resource-efficient profile also performed well, especially under medium competition. In contrast, the performance-centric profile had the lowest energy savings, indicating that prioritizing processing speed alone is less effective. The general profile exhibited lower optimization performance, attributable to its balanced yet less energy-oriented strategy. A detailed comparison is presented in Table VI, with some individual values rounded for clarity.

### B. Analysis of Scheduling Profiles

The energy-centric profile consistently outperformed other strategies, achieving energy optimization of 37.96% in low competition, 39.13% in medium, and 33.82% in high competition. The resource-efficient profile also performed well in low (26.80%) and medium (32.70%) competition but showed a significant drop to 4.86% in high competition, indicating its limitation under heavy resource contention.

TABLE VI. ENERGY CONSUMPTION: LOW COMPETITION

| Profile | Default K8s (kJ) | TOPSIS (kJ) | ≈ Energy Savings (kJ) | Optimization (%) |
|---|---|---|---|---|
| **Low Competition** | | | | |
| General (Balanced) | 0.5036 | 0.4586 | 0.0450 | 8.93 ▼ |
| Energy-centric | 0.5036 | 0.3124 | 0.1912 | 37.96 ▼ |
| Performance-centric | 0.5036 | 0.4924 | 0.0112 | 2.22 ▼ |
| Resource-efficient | 0.5036 | 0.3686 | 0.1350 | 26.80 ▼ |
| Average (Low) | 0.5036 | 0.4080 | 0.0956 | **18.98** ▼ |
| **Medium Competition** | | | | |
| General (Balanced) | 0.4375 | 0.3650 | 0.0725 | 16.57 ▼ |
| Energy-centric | 0.4375 | 0.2663 | 0.1712 | 39.13 ▼ |
| Performance-centric | 0.4375 | 0.4037 | 0.0338 | 7.72 ▼ |
| Resource-efficient | 0.4375 | 0.2944 | 0.1431 | 32.70 ▼ |
| Average (Medium) | 0.4375 | 0.3324 | 0.1052 | **24.03** ▼ |
| **High Competition** | | | | |
| General (Balanced) | 0.4471 | 0.3867 | 0.0604 | 13.50 ▼ |
| Energy-centric | 0.4257 | 0.2817 | 0.1440 | 33.82 ▼ |
| Performance-centric | 0.4257 | 0.3904 | 0.0353 | 8.29 ▼ |
| Resource-efficient | 0.4257 | 0.4050 | 0.0207 | 4.86 ▼ |
| Average (High) | 0.4311 | 0.3660 | 0.0651 | **15.12** ▼ |
| Average (All) | 0.4574 | 0.3688 | 0.0886 | **19.38** ▼ |

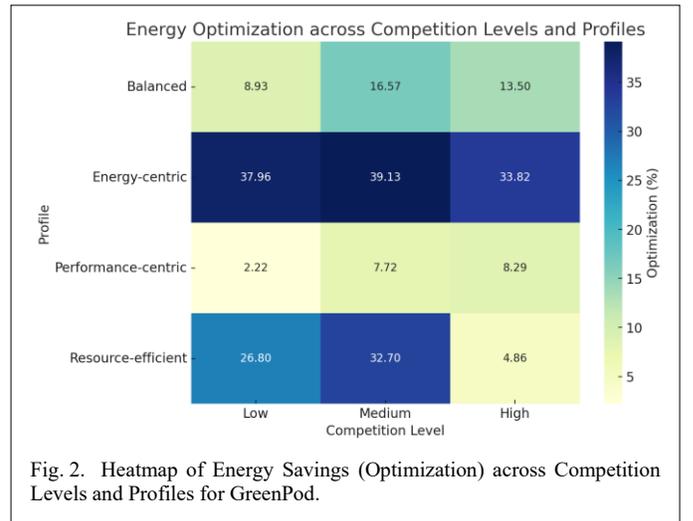

Fig. 2. Heatmap of Energy Savings (Optimization) across Competition Levels and Profiles for GreenPod.

The performance-centric profile exhibited the lowest energy optimization across all levels (2.22%, 7.72%, and 8.29%), demonstrating that prioritizing execution speed without considering energy efficiency leads to suboptimal results. Notably, this profile was the only one that improved as competition increased, suggesting a unique adaptive response to higher system loads, possibly due to more efficient utilization of computational resources when the system is more fully loaded.

The general (balanced) profile maintained uniform resource distribution but demonstrated lower energy optimization (8.93%, 16.57%, and 13.50% across competition levels), reflecting the trade-off between consistency and energy savings. Figure 2 visualizes the energy optimization achieved by each scheduling strategy across different competition levels.

### C. Impact of Competition Levels

Figure 2 clearly shows that the energy-centric approach consistently outperforms the default Kubernetes scheduler at all competition levels. In contrast, the performance-centric profile, indicated by lighter shades, shows minimal optimization, especially in low-competition scenarios. This highlights that prioritizing performance metrics alone does not lead to energy savings when system load is low.

Our analysis also reveals that competition level significantly influences scheduling effectiveness. Medium competition consistently provides the optimal conditions for energy savings, with an average optimization of 24.03% across all profiles. Low competition environments yield moderate optimization potential (18.98% average), while high competition presents the most challenging operational conditions (15.12% average). These findings suggest that scheduling strategies should adapt dynamically to system load. In low and medium competition environments, energy-centric strategies are preferable, while high competition may require hybrid approaches balancing energy awareness with resource efficiency.

### D. Node Allocation and Workload Analysis

Energy-centric strategies tend to allocate workloads to energy-efficient nodes (Category A), minimizing energy consumption. In contrast, performance-centric strategies distribute workloads across high-capacity nodes, leading to higher energy usage without proportional performance gains.

Furthermore, energy-centric scheduling is particularly effective for computationally intensive workloads, with medium workloads showing the highest savings. Light workloads, however, exhibit variable results due to scheduling overhead, indicating that energy-centric strategies work best for demanding tasks (e.g., machine learning or edge AI-related tasks).

Moreover, in high-competition environments, the energy-centric profile remains effective, though its efficiency decreases as resource utilization nears capacity. Combining energy-centric strategies with dynamic load balancing could further improve performance in such scenarios.

### E. Real-World Impact Analysis

To assess the broader impact of our findings, we extrapolated potential energy savings to real-world environments using operational data from the SURF Lisa Compute cluster as a benchmark. Our extrapolation methodology leverages empirical job statistics derived from SLURM scheduler logs analyzed by Chu et al. [31]. Between January 2022 and January 2023, the SURF Lisa Compute cluster processed an average of 6,304 jobs daily, with peak loads reaching 163,786 jobs. The workload composition comprised 13.32% machine learning tasks and 86.68% generic computational jobs, reflecting the diverse workload distribution typical of high-performance computing environments.

Assuming containerized job deployment and applying our GreenPod energy-centric scheduling approach with an average optimization of 19.38% across all competition levels, we conducted an energy impact assessment for a comparable environment. The average job energy consumption was calculated as 0.024 kWh, based on the power model for blade servers proposed by Dayarathna et al. [32]: $P\_blade = 14.45 + 0.236u\_cpu - (4.47E-8)u\_mem + 0.00281u\_disk + (3.1E-8)u\_net$ watts [32]. Using typical workload parameters (60% CPU utilization, 8M memory accesses/sec, 350 I/O ops/sec, 3M network ops/sec) with a 34-minute average runtime and PUE of 1.45, we derived the 0.024 kWh consumption [32]. Using these parameters, implementing our scheduling optimization would yield an estimated daily energy savings of approximately 0.0293 MWh (0.024 kWh × 6,304 jobs × 0.1938). This equates to cumulative savings of about 0.88 MWh monthly and 10.70 MWh annually for a single cluster deployment.

Extending this analysis to a medium-sized data center comprising around 10 similar clusters (processing approximately 63,040 jobs daily), the potential energy impact scales proportionally. In this setting, GreenPod could achieve energy savings of 0.293 MWh per day, 8.80 MWh per month, and approximately 107.02 MWh annually. These results demonstrate GreenPod's contribution to both operational cost reduction and environmental sustainability in large-scale computing infrastructures.

*F. Environmental and Economic Benefits*

GreenPod's energy-centric scheduling approach delivers substantial environmental and economic benefits at scale. To quantify these impacts, we further conducted a comprehensive analysis using established conversion factors and market valuations.

**Environmental Impact**: Based on the annual energy savings calculated earlier (10.70 MWh per cluster), we estimated the corresponding $CO_2$ emission reductions. According to the EPA's Emissions & Generation Resource Integrated Database (eGRID), the U.S. national average emission factor is approximately 0.823 pounds of $CO_2$ per kWh [33]. To convert this to metric units, we multiply by 0.4536 kg/lb and 1,000 kWh/MWh, yielding approximately 373.2 kg $CO_2$ per MWh. Applying this factor, the annual reduction in $CO_2$ emissions from a single cluster amounts to approximately 3.99 metric tons (10.6872 MWh × 373.2 kg $CO_2$/MWh). That is, the annual reduction in $CO_2$ emissions would be approximately 3.99 metric tons for a single cluster as that of the SURF Lisa Compute-scale cluster and 39.94 metric tons for a medium-sized data center comprising 10 clusters.

According to the EPA's Greenhouse Gas Equivalencies Calculator (2022), this reduction is equivalent to removing approximately 0.87 passenger vehicles from the road for one year for a single cluster, or 8.70 vehicles for a medium-sized data center with 10 clusters, based on the average passenger vehicle emitting 4.6 metric tons of $CO_2$ per year [34].

**Economic Impact**: Translating these savings into financial terms, based on an average commercial electricity rate of $0.1289 per kWh (as reported by the U.S. Energy Information Administration, 2025) [35], a SURF Lisa-scale cluster would save approximately $1,380 annually in direct electricity costs. A medium-sized data center with 10 similar clusters would save approximately $13,795 annually.

**Additional Economic Considerations:** Additionally, the value derived from carbon credits can vary significantly based on the pricing mechanism and region. According to the World Bank Carbon Pricing Dashboard (2024) [36], carbon credit prices range from $0.46 to $167 per metric ton of $CO_2$. Using this range, the potential annual value of carbon credits for a single cluster would be between $1.84 (3.99 metric tons × $0.46) and $667 (3.99 metric tons × $167). For a medium-sized data center with 10 clusters, the annual credit value would range from $18.40 to $6,670.

**Combined Financial Impact:** Combining direct energy savings and carbon credit value, the total annual financial benefit per SURF Lisa-scale cluster can range from approximately $1,380 to $2,047, while a medium-sized data center with 10 clusters would see savings ranging from $13,814 to $20,465. Over a standard five-year planning period, this amounts to $6,907 to $10,233 for one cluster and $69,068 to $102,326 for a data center with 10 clusters.

TABLE VII. ENERGY AND COST SAVINGS ASSESSMENT

| Metric | Single Cluster (e.g., SURF Lisa) | Medium-Sized D.C. (10 Clusters) |
|---|---|---|
| Daily Energy Savings | 0.0293 MWh | 0.29 MWh |
| Monthly Energy Savings | 0.88 MWh | 8.80 MWh |
| Annual Energy Savings | 10.70 MWh | 107.02 MWh |
| Annual $CO_2$ Reduction | 3.99 metric tons | 39.94 metric tons |
| Vehicles Removed | 0.87 vehicles | 8.70 vehicles |
| Annual Cost Savings | $1,380 | $13,795 |
| Total Savings (1 Yr, Min) | $1,381 | $13,814 |
| Total Savings (1 Yr, Max) | $2,047 | $20,465 |
| Total Savings (5 Yrs, Min) | $6,907 | $69,068 |
| Total Savings (5 Yrs, Max) | $10,233 | $102,326 |

These assessments, summarized in Table VII, demonstrate that GreenPod offers environmental sustainability along with economic advantages. While the financial impact is more modest than initially calculated, integrating energy-efficient scheduling across large-scale computing infrastructures like the SURF Lisa cluster can still contribute to reducing operational costs and carbon footprints.

## VI. CONCLUSION

GreenPod's TOPSIS-based Kubernetes scheduler reduces energy consumption by up to 39.1% compared to the default scheduler, particularly in medium-complexity, multi-threaded inference tasks. Utilizing five weighted criteria for smart pod placement, it performs well in medium competition environments with minimal scheduling overhead. Implementing GreenPod at scale yields modest economic benefits, with a single SURF Lisa-scale cluster saving about $1,380 annually and a medium-sized data center saving $13,795 per year. Although lower than expected, these savings support the business case for adoption, factoring in direct energy reductions and carbon credit potential. Our results demonstrate that energy-centric scheduling is a viable strategy for sustainable container management. For future work, we plan to enhance the efficiency of GreenPod for lightweight tasks, employ adaptive profiling through machine learning, and develop hybrid approaches for high-competition scenarios.